\documentclass{appolb}
\usepackage[pdftex]{graphicx}
\pdfoutput=1

\begin{document}
\title{Applications of neutron activation spectroscopy%
\thanks{Presented at Symposium on applied nuclear physics and innovative technologies, Cracow, 03-06 June 2013.}%
}
\author{M. Silarski for the SABAT collaboration\footnote{The SABAT Collabotation: P. Moskal, M. Silarski, M. Smolis, S. Tadeja}
\address{Institute of Physics, Jagiellonian University, PL-30-059 Cracow, Poland}}
\maketitle
\begin{abstract}
Since the discovery in 1932, neutrons became a basis of many methods used not only in research, but also in industry
and engineering. Among others, the exceptional role in the modern nuclear engineering is played by the neutron
activation spectroscopy, based on the interaction of neutron flux with atomic nuclei. In this article we shortly
describe application of this method in medicine and detection of hazardous substances.  
\end{abstract}
  
\section{Introduction}
\begin{figure}[htb]
\centerline{%
\includegraphics[width=10.cm]{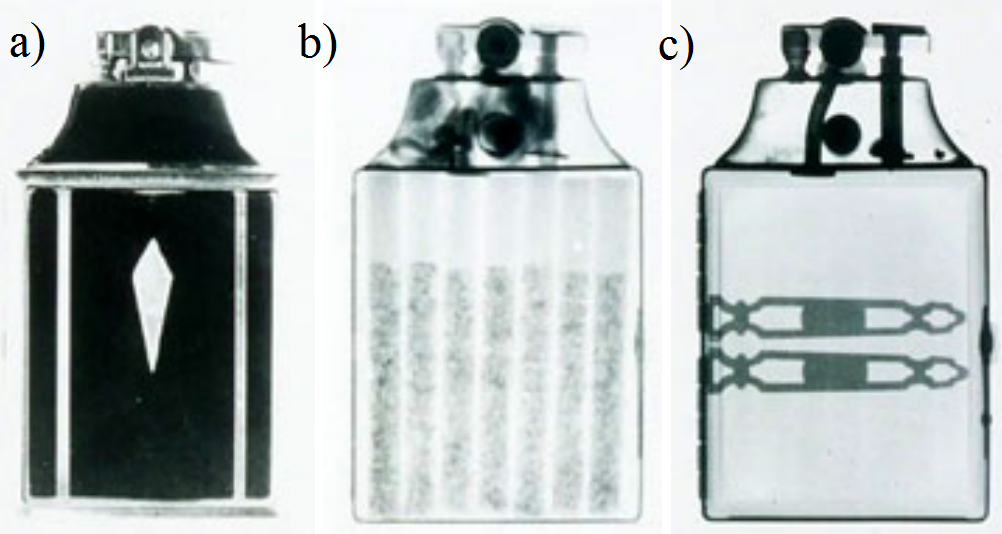}}
\caption{Comparative radiographs of a cigarette case: a) photograph, b) neutron radiography, c) X-rays radiography.
The figure is adapted from~\cite{mnrc}.}
\label{fig:1}
\end{figure}

Neutrons have many properties that make them ideal for certain types of application, especially concerning imaging
and elemental traces detection techniques. As neutral particles neutrons can deeply penetrate even very dense matter.
On the other hand they interact strongly with nuclei, which is a great advantage over eg. X-rays which
interact almost only with electrons in the matter, and thus are "blind" to the type of element.\\
One of the most important technique being used widely in research and engineering for a long time is the neutron radiography.
As the name implies, a neutron beam penetrates the object to be studied. This beam is attenuated by the sample material
depending on the neutron interaction cross-sections. The beam is then detected by a two-dimensional imaging device that
provides an image representing the macroscopic structure of the studied sample~\cite{mnrc,nares}. An example of radiographs
taken with different methods is presented in Fig.~\ref{fig:1}, where one can clearly see the differences between X-rays
and neutrons. 
Analogously as for X-rays, a series of n-ray radiographs provides a three-dimensional image of the object inner structure,
i.e. its tomographic picture. The sensitivity to light elements, especially to Hydrogen, makes the n-ray radiography a very
valuable alternative to the X-rays. It is used for example in studying the flow of oil in automobile transmissions, monitoring
complex piping systems and fluid flow, to determine the pyrotechnic product quality and in many other applications~\cite{mnrc}.\\
Passing through matter neutrons interact with nuclei in several processes including the elastic and inelastic scattering,
neutron capture, fission or particle emission from the nucleus. The inelastic scattering and neutron capture constitute
base of the Neutron Activation Analysis techniques (NAA), which are widely used for the quantitative elemental concentrations
determination. These two types of neutron interaction results in the nuclei excitation turning them into radioactive nuclides,
which decay emitting gamma rays. Their energies are characteristic for each element, so that the measurement of these gamma
rays intensities gives information on the elemental content of the irradiated object.
Methods based on NAA have a big application potential since they provide an analysis of a large number of elements simultaneously
and non-destructively. Moreover, they have a very low detection limits for many elements. Devices based on NAA are used widely
in industry (eg. crude oil and coal deposits identification, water monitoring), medicine and homeland security.
The latter two applications will be discussed in more details in next sections of this contribution. 
\section{Medicine: breast cancer diagnostics}
The presently used breast cancer diagnostic methods are based mainly on X-rays imaging (mammography), and have reached
almost 98\%  sensitivity in detecting tumors~\cite{calsec1}. However, mammography do not provide sufficiently accurate
information which would allow to determine whether the tumor is benign or malicious. Moreover, for confirmation of the
diagnosis one needs invasive biopsy.
The other disadvantage is, that the radiologists working on the interpretation of the mammography images need a long and
expensive training before they are able to set a good diagnosis.\\
Several studies have demonstrated that breast cancer is associated with changes in the concentration of some elements
(eg. calcium, potassium or iron) in the malignant tissue at very early stages\footnote{Changes of trace element
concentrations in human tissue may be a precursor to malignancy in several other organs like brain or
prostate~\cite{Kapadia}}. These changes often occur much earlier than morphologic changes such as tumors and calcifications.
Thus, detection of these changes gives a chance to diagnose of breast cancer much
before the tumor grows large enough to be detected by existing imaging techniques~\cite{Kapadia}. The NAA based techniques 
are ideal to quantify the concentration of elements of interest in the tissue providing at the same time location of region
affected by the disorder and diagnosis~\cite{Kapadia}. Similar method, based on oxygen concentration measurement,
is developed by CALSEC~\cite{calsec1}. It makes use of the  fact that cancerous tumors chemically differ from 
healthy tissue by the oxygen content. Therefore, any oxygen concentration difference between a tumor and the adjacent
healthy tissue indicates that it is malicious. This method is intended to support the conventional methods providing
kind of "needleless biopsy".\\
Neutron based techniques for cancer diagnosis are at present at an advanced stage of development and will be soon
a very useful alternative or supplement of mammography.
\section{Homeland security: detection of hazardous substances}
\begin{figure}[htb]
\centerline{%
\includegraphics[width=15.cm]{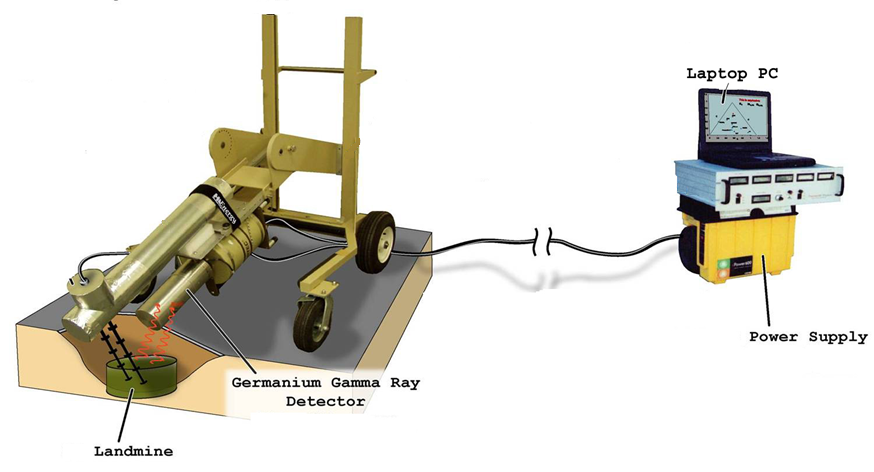}}
\caption{Illustration of a system for hazardous substances detection based on neutrons.
The figure is adapted from~\cite{maglich_ppt}.}
\label{fig:2}
\end{figure}

In the face of increasing danger of terrorist attacks and still growing trade of illicit drugs, a development of new
hazardous substances detection technologies became of a great importance. At present, the detection of concealed
contraband is based mainly on the X-rays, vapour detection and sniffer dogs. X-ray scanners suffer from
the main disadvantage of having a small cross section interaction with low electron density elements from
which organic materials, including most explosives and illicit drugs, are composed. Thus, they are detected 
mainly through shape recognition, which could be very difficult since the hazardous material can be molded or
packed into any form~\cite{Buffler}. In fact, up to now the false alarm rates in the airport luggage inspections
amount almost to 100\%.
Also the vapour detection and sniffer dog methods suffer from a high rate of false alarms, especially in the case
of people professionally working with explosives or drugs. These methods can be used only for security purposes
at airports, in the Civil Service etc. In the military application the only devices used for example for
area demining are metal detectors. In the ground full of metal remains this process became very expensive and time
consuming. Moreover, the novel landmines are being made without any metal parts, thus they cannot be detected with
those devices.\\
Neutron-based techniques were considered to be used in the hazardous substances detection for more than forty years,
but the first functioning devices were designed and produced at the beginning of XXI-th century by B. Maglich~\cite{maglich}.
They utilize the fact, that most of the commonly used explosives, drugs or war gases are organic materials. Therefore,
they are composed mostly of oxygen, carbon, hydrogen and nitrogen (see Tab.~\ref{tab:1}).  
\begin{table}
\begin{center}
\begin{tabular}{|c|c|c|}
\hline
Substance & Stoichiometric formula & Ratio C:H:N:O\\
\hline
Trinitrotoluene (TNT) & C$_7$H$_5$N$_3$O$_6$ & 1.2 : 0.8 : 0.5 : 1 \\
\hline
Hexogen (RDX) & C$_3$H$_6$N$_6$O$_6$ & 0.5 : 1 : 1 : 1\\
\hline
Nitroglycerine & C$_3$H$_5$N$_3$O$_9$ & 0.33 : 0.56 : 0.33 : 1\\
\hline
Cocaine & C$_{17}$H$_{21}$N$$O$_4$ & 4.25 : 5.25 : 0.25 : 1\\
\hline
Heroine & C$_{21}$H$_{23}$N$$O$_5$ & 4.2 : 4.6 : 0.2 : 1\\
\hline
Amphetamine & C$_9$H$_{21}$N$$O$_4$ & 4.25 : 5.25 : 0.25 : 1\\
\hline
\end{tabular}
\end{center}
\caption{
\label{tab:1}
Molecular content of few chosen hazardous substances.}
\end{table}
Thus, these substances can be unambiguously identified by the determination of the ratio
between number of C, H, N and O atoms in a molecule, which can be done noninvasively applying
Neutron Activation Analysis techniques. The schematic illustration of one of the possible practical
realizations of a device utilizing this method is shown in Fig~\ref{fig:2}.
The suspected item can be irradiated with a flux of neutrons produced using compact generators based
e.g. on deuteron-tritium fusion, where deuterons are accelerated to the energy of 0.1 MeV and hit a solid
target containing tritium. As a result of the fusion an alpha particle is created together with the neutron,
which is emitted nearly isotropically with a well defined energy equal to about 14.1 MeV.
Such energy is sufficient to excite nuclei of carbon $^{12}$C (4.43 MeV), oxygen $^{16}$O (6.13 MeV)
and nitrogen $^{14}$N (2.31 MeV and 5.11 MeV)~\cite{moskalAnn}. $\gamma$ quanta from  the de-exciting nuclei
are then detected by a germanium detector providing a very good energy resolution. Counting the number
of gamma quanta corresponding to a given nuclei provides information about the stoichiometry of
the examined item. Since the hydrogen nuclei cannot be excited by neutron inelastic scattering the amount
of hydrogen atoms in the sample cannot be examined using this method. Yet, hazardous substances can be
discriminated from non-hazardous by the atomic ratio of carbon, oxygen and nitrogen only.
There are two major difficulties which has to be solved in this type of detectors.
One is an isotropic generation of neutrons which induces unfortunately a large background from irradiation
of material surrounding the interrogated object. This noise can be  significantly reduced by the requirement
of the coincident detection of the alpha particle, which allows for the neutron tagging by registration
of the direction of alpha particle\cite{moskalAnn}. Moreover, germanium detectors are quite expensive
and need a cooling system which significantly limit the mobility of the scanner. Thus, there has been done
a lot of studies to replace germanium with detector systems based on scintillators. 
\section{Summary}
Over the last decade, technologies using fast neutrons have found use in a variety of contexts
in the research, industry and engineering. Especially the Neutron Activation Analysis based techniques
were found to be very  promising. They may find applications e.g. in remote and non -invasive
detection of explosives or drugs, as well as in medical diagnostics of cancer tissues. Thus, they could
open a new frontiers in homeland security and biomedical sciences.
\section*{Acknowledgements}
This work was supported by the Polish Ministry of Science and Higher Education through grant No. K/DSC/000681
for
the support of young researchers and PhD students of the Department of Physics, Astronomy and Applied
Computer Science of the Jagiellonian University.

\end{document}